\documentclass[12pt]{article}
\usepackage{amsmath}
\usepackage{graphicx}
\usepackage{enumerate}
\usepackage{natbib}
\usepackage{url} 
\usepackage{microtype}
\usepackage[utf8]{inputenc} 
\usepackage[T1]{fontenc}    
\usepackage{hyperref}       
\usepackage{url}            
\usepackage{booktabs}       
\usepackage{amsfonts}       
\usepackage{nicefrac}       
\usepackage{microtype}      
\usepackage{lipsum}		
\usepackage{graphicx}
\usepackage{natbib}
\usepackage{doi}
\usepackage{bbm}
\usepackage{dirtytalk}
\usepackage{amsmath}
\usepackage{array}
\usepackage{lineno}
\usepackage{booktabs}
\usepackage{tabularx}
\usepackage{algorithm}
\usepackage{algpseudocode}

\newcommand{\blind}{1}

\begin{document}

\bibliographystyle{plainnat}

\def\spacingset#1{\renewcommand{\baselinestretch}%
{#1}\small\normalsize} \spacingset{1}


\if1\blind
{
  \title{\bf Discovery of Critical Thresholds in Mixed Exposures and Estimation of Policy Intervention Effects}
  \author{David B. McCoy \thanks{
    Corresponding author: david\_mccoy@berkeley.edu. The authors gratefully acknowledge funding for Core E of the NIEHS Superfund Center at Berkeley funded by NIH grant P42ES004705}\hspace{.2cm}\\
    Division of Biostatistics, University of California, Berkeley\\
    and \\
    Alan Hubbard \\
    Division of Biostatistics, University of California, Berkeley \\ 
    and \\ 
    Mark van der Laan \\ 
    Division of Biostatistics, University of California, Berkeley \\ 
    and \\
    Alejandro Schuler \\ 
    Division of Biostatistics, University of California, Berkeley \\ }
  \maketitle
} \fi

\if0\blind
{
  \bigskip
  \bigskip
  \bigskip
  \begin{center}
\setlength{\baselineskip}{1.5\baselineskip}
    {\LARGE\bf Discovery of Critical Thresholds in Mixed Exposures and Estimation of Policy Intervention Effects}
\end{center}
  \medskip
} \fi

\bigskip
\begin{abstract}

Traditional regulations of chemical exposure tend to focus on single exposures, overlooking the potential amplified toxicity due to multiple concurrent exposures. We are interested in understanding the average outcome if exposures were limited to fall under a multivariate threshold. Because threshold levels are often unknown \textit{a priori}, we provide an algorithm that finds exposure threshold levels where the expected outcome is maximized or minimized. Because both identifying thresholds and estimating policy effects on the same data would lead to overfitting bias, we also provide a data-adaptive estimation framework, which allows for both threshold discovery and policy estimation. Simulation studies show asymptotic convergence to the optimal exposure region and to the true effect of an intervention. We demonstrate how our method identifies true interactions in a public synthetic mixture data set. Finally, we applied our method to NHANES data to discover metal exposures that have the most harmful effects on telomere length. We provide an implementation in the open-source CVtreeMLE R package. 

\end{abstract}

\noindent%
\vfill

\newpage
\spacingset{1.9} 
\section{Introduction}
\label{sec:intro}

Policymakers and health researchers seek to understand how joint exposures impact health outcomes and to find and establish safe exposure thresholds based on concurrent exposure. Traditionally, regulatory frameworks have been oriented towards setting exposure limits based on individual chemicals. However, in the complex environment of real-world settings, individuals are invariably exposed to a plethora of chemicals simultaneously, often referred to as mixed exposures. Such exposures range from air pollutants and endocrine disruptors to pesticides and heavy metals. \cite{Landrigan2006, Vrijheid2009, Kortenkamp2014, Braun2018, NationalResearchCouncil2018}. 

The complexity of mixed exposures arises when the joint impact of such chemical mixtures deviates from the additive effects of their individual components. Combined exposures could potentially increase health risks beyond the sum of the effects of individual agents \cite{nrc2008}. In contrast, some exposures might counteract the impacts of others. Furthermore, the effects of these combined exposures can vary due to environmental stressors, genetic predispositions, and psychosocial determinants \cite{carlin2013}. Given this complexity, there is an urgent call for cutting-edge, assumption-lean statistical methodologies suitable for delimitating exposure thresholds for concurrent exposure \cite{Safe1993, Kortenkamp2007} to help inform safe exposure limits. 

In modern causal inference concerning mixed exposures, conventional approaches have relied heavily on parametric models. More often than not, consideration is given to the effect of a singular exposure, typically represented as a coefficient within a generalized linear regression model (GLM). Such piecemeal evaluations fail to capture the collective effect of multiple concurrent exposures, a common scenario in realistic exposure settings \cite{Davidson2009, Kim2014, Eze2011}. Although efforts have been made to estimate the joint effects of mixed exposures - examples including the popular Weighted Quantile Sum Regression (WQS) \cite{Keil2019}, Bayesian Mixture Modeling \cite{DeVocht2012} and Bayesian Kernel Machine Regression \cite{Bobb2014} - they come with their caveats. Some of these methods harbor strong inherent assumptions, such as the directional homogeneity and linear/additive assumptions in WQS. Bayesian approaches demand prior knowledge or risk human-induced bias due to subjective choice of priors. Flexible methods such as BKMR are often visualized through a variety of dose-response graphs rather than distilled into an intuitive summary statistic \cite{Bobb2014}. To date, none of the methods have been built to efficiently estimate a predefined target parameter, starting from a full-causal model and explicitly stating the assumptions necessary to identify the parameter in the observed data \cite{van2011targeted, van2018targeted}. We attempt to bridge this gap. 

Ideally, the analyst would have the full multidimensional dose-response relationship, but this is complex due to unrealistic assumptions and nonpathwise differentiability, complicating robust confidence interval construction. To address this, we suggest categorizing joint exposures to assess the impact of specific exposure thresholds on average outcomes.  If our aim is to quantify the effect of imposing certain exposure thresholds, a sensible approach is to categorize the joint exposure and compare the average outcome if everyone were forced to have exposures within a region compared to the observed average outcome under observed exposures. Because this simulated policy is based on a region discovered in the exposure space, we call our target parameter the attributable regional exposure effect (ARE). This approach is helpful because interpretable categories can be defined, such as $(A_1 < a_1) \& (A_2 < a_2)$, where $a_i$ are specific values in $A$ that are of clear interest to policy makers. Identification assumptions are also more transparent in this setting. 
In this context, we seek to categorize the exposure space and find the regions with the expected minimum (or maximum) population outcome. However, this intervention that effectively forces people to a region is ambiguous. The post-intervention probability density of exposure $A$ is concentrated in the given region, but there are infinitely many different policies that achieve this. Our choice preserves relative natural selection preferences within the region, per covariate stratum, which is a natural option and lends itself to easy estimation with standard tools. However, the thresholding of the exposure space that minimizes/maximizes the population average outcome is not known a priori. Data must be used to determine which regions are relevant. To estimate this oracle region of the exposure space, we develop a recursive partitioning algorithm paired with targeted learning to greedily search the exposure space using a fast machine learning algorithm for propensity score and outcome regression. Once the thresholds have been identified, we obtain more precise inference for the effects using a cross-validated targeted maximum likelihood estimation (CV-TMLE), with a larger set of algorithms in the ensembles used to estimate the propensity and outcome regression. 

Using the same data to define thresholds and estimate effects in the outcome minimum region can lead to biased results. We address this problem by splitting the data, estimating thresholds on one part, and estimating counterfactual effects (given the fixed thresholds) on the other. To efficiently use all the data, we repeat this splitting process in a round-robin fashion (K-fold cross-validation). We build this framework on previous work related to data-adaptive parameters \cite{Hubbard2016} and cross-validated targeted minimum loss-based estimation (CV-TMLE) \cite{Zheng2010}, our method, called \texttt{CVtreeMLE}, is a novel approach for estimating the joint impact of a mixed exposure by using CV-TMLE that guarantees consistency, efficiency, and multiple robustness despite using highly flexible learners (ensemble machine learning) to estimate a data-adaptive parameter. A crucial challenge in this process is the inconsistency of the data-adaptive parameter. If the region found to minimize/maximize the expected outcome is not the same across all folds the interpretability of the pooled parameter is difficult, although fold specific results still allow for inference \cite{HubbardKennedyVanDerLaan2018}. 

We provide implementations of this methodology in our free and open-source software package CVtreeMLE, written in the R language and environment for statistical computing (R Core Team, 2022). The CVtreeMLE software package has undergone a rigorous peer review process by the Journal of Open Source Software, which validates the robust implementation of our innovative methodology for the analysis of mixed exposure \cite{McCoy2023}.

This manuscript is structured as follows: Section 2 discusses semiparametric methodology and redefines the population intervention effect as the attributable regional effect (ARE). Section 3 focuses on ARE estimation and inference, while Section 4 introduces the oracle target parameter. Section 5 delves into cross-estimation, including pooling methods. Simulation studies demonstrating the estimator's properties are in Section 6. Section 7 applies CVtreeMLE to NIEHS data and compares it with other methods, section 7.2 we apply our method to NHANES data. The final sections (8 and 9) detail the \texttt{CVtreeMLE} software and conclude with a discussion.

\section{The Estimation Problem}

\subsection{Setup and Notation} 
We consider an observational study with baseline covariates ($W \in \mathbb{R}^p$), multiple exposures ($A \in \mathbb{R}^m$), and a single timepoint outcome ($Y$). Let $O = (W,A,Y)$ denote the observable data. We assume that there exists a potential outcome function $Y(a)$ (that is, $Y(a)$ is a random variable for each value of $a$) that generates the outcome that would have been obtained for each observation had exposure been forced to the value $A=a$. These potential outcomes are not observed, but the observed outcome $Y$ corresponds to the potential outcome for the observed value $A$ of the exposure, that is, $Y=Y(A)$. We use $P_0$ to denote the true data-generating distribution. We assume that our $O_1, O_2,...,O_n$ are independent and identically distributed (i.i.d.) draws of $O = (W,A,Y) \sim P_0$. We decompose the joint density as $p_{Y,A,W}(y,a,w) = p_{Y|A,W}(y,a,w)p_{A|W}(a,w)p_W(w)$ and make no assumptions about the forms of these densities. Let $\mu(a,w) = E[Y|A=a, W=w]$ represent the observable (stratum-specific) dose-response curve that can actually be estimated from observable data. 

Let $\Psi$ map from the set of possible data-generating distributions to the real numbers. Let $\Psi(P_0)$, the value of this map at the true, unknown distribution $P_0$, denote a target parameter (estimand) of interest. We can view our observed data $(O_1 \dots O_n)$ as a (random) probability distribution $P_n$ that assigns a probability mass $1/n$ to each observation $O_i$. Let $\hat\Psi$ mapping from empirical distributions to numbers denote our estimator (algorithm) such that $\hat\Psi(P_n)$ is our estimate from data.

\subsection{Defining the Differential Effect Given Regional Exposure}

We ask about the differential policy effect of allowing self-selection within a region of the exposure space. Let us consider a situation where there is a community with houses and families have certain probabilities of moving to live in each house. A state enacts a law that restricts housing options to those more than 5 km from a chemical plant. In some situations, such as housing vouchers, it may make sense to say that the new distribution of housing choices under restriction is simply a scaled-up version of the density with no restriction. In this case, the preference profile of houses in the allowed region may change proportionally compared to the unrestricted distributions. This assumption is only realistic in certain circumstances, although it is implicit when we binarize any exposure data. \cite{lee2024bridging}

For example, if a family were twice as likely to choose House 1 as they were to choose House 2, that relative preference would continue to hold regardless of whether House 3 was an option. But, of course, this may not always be the case and preferences for House 1 and 2 could be different under the House 3 restriction. This is a special kind of modified treatment policy. Formally, let $\mathcal{A} \subset \mathbb{R}^m$ be a subset of the exposure space. For example, assume that $\mathcal{A}$ represents allowable neighborhoods or doses of drugs/pollutants that have been deemed safe for combination. Let $T$ denote the binary random variable $1_{\mathcal{A}}(A)$ indicating whether an observation complied with the policy. Let $\pi_{\mathcal{A}}(w) = P(A \in \mathcal{A} \mid W=w)$ be the probability that an observation with covariates $w$ naturally self-selects treatment in the requested region $\mathcal{A}$. We define the modified treatment variable $\tilde{A}$ to represent the exposure distribution once all exposures are forced to self-select within the region $\mathcal{A}$. The modified treatment has the following density:
\[
p_{\tilde{A} \mid W}(a, w) = \frac{1_{\mathcal{A}}(a)}{\pi_{\mathcal{A}}(w)} p_{A \mid W}(a, w),
\]
which preserves the relative self-selection preferences for each available exposure and sets the preferences for "outlawed" exposures to zero. Now we can define the expected population outcome if we were to impose this policy:
\begin{align*}
E[\mu(\tilde{A}, W)] &= \int \mu(a, w)\, dP_{\tilde{A}, W} \\
&= \int \mu(a, w) \frac{1_{\mathcal{A}}(a)}{\pi_{\mathcal{A}}(w)}\, p_{A \mid W}(a, w) p_{W}(w)\, da\, dw \\
&= \int_w \left[ \int_{a \in \mathcal{A}} \mu(a, w) \frac{p_{A \mid W}(a, w)}{\pi_{\mathcal{A}}(w)} \, da \right] p_{W}(w) \, dw \\
&= E[E[Y \mid T =1, W=w]].
\end{align*}
What we have shown is that this parameter is a population average of some function $Q(\mathcal{A}, W)= E[Y \mid \mathcal{A}, W]$ when $\mathcal{A}$ is forced to 1. For any value of $w$, $Q(1, w)$ is a particular convexly weighted average of the causal dose-response curve across the different exposure levels, thus collapsing it down to a single number for each $w$. We want to compare the average outcome if all individuals had exposures within the region $\mathcal{A}$ with the average observed outcome under the observed exposure distribution. This defines our parameter of interest, which we call the average regional effect (ARE): $\psi = E[Q(T = 1, W)] - E[Y]$, representing the difference in the average outcomes if we forced exposure to self-selection within $\mathcal{A}$ versus the observed outcomes.

If we do not assume relative self-selection, we would need to account for how each individual's preference might change under the restriction. For any regulation or restriction, we would need to model the conditional density of exposures in the region given covariates to try and estimate how the exposure density profile would change under restriction. Alternatively, we would need to make assumptions about what the distribution would look like under restriction, for example, perhaps aggregating just below the threshold, leading to higher densities near the boundary and lower densities farther from the threshold. However, we do not know this information a priori. Therefore, in practice, this approach would involve building a density estimator for $p(A \mid W, A \in \mathcal{A})$, which accounts for the restricted set of options for all possible policies, which is computationally infeasible. We would then use these individualized densities to calculate the new exposure levels for each person under the restriction, which would require additional data and assumptions about how preferences change.

For reasons of interpretability, frequency of binarizing exposures in the literature (with aforementioned implicit assumptions), and ease of using existing tools for estimation, we choose to estimate a policy under relative self-selection. 

In most applications, it is not known a priori which region $\mathcal{A}$ should be set. For example, we may not know how various chemicals or drugs interact and how to set safe limits for all of them. Therefore, $\mathcal{A}$ itself is, in practice, something that should be estimated to maximize some objective. However, for the purposes of establishing our theory, we should first imagine $\mathcal{A}$ as known and fixed. We will then show how we can choose what policy to enact (i.e., choose $\mathcal{A}$) while also unbiasedly estimating its effect. We first discuss the assumptions necessary for the statistical quantity estimable from the sample data to have causal interpretations, then we move on to discuss how $\mathcal{A}$ can be determined.

\subsection{Identification and Causal Assumptions}

Our target parameter is defined in terms of the observable distribution and does not represent a causal quantity without additional assumptions. To add a causal interpretation, we define the \textit{causal} ARE as $\psi^* = E[Y(A)|T = 1] - E[Y]$. Standard conditioning arguments show $\psi^* = \psi$ when the conditional randomization assumption, $A \perp Y(a) \mid W$ for all $a \in \mathcal A$ and positivity: $P(T = 1 | W) > 0$ for all $w$, are met. The randomization assumption, asserts that the potential outcomes are independent of exposure within covariate-defined strata, indicating that there are no unmeasured confounders. The positivity assumption, $P(T = 1 | W) > 0$ for all $w$, ensures that all exposure levels and covariate strata are represented in the data, critical for accurate effect estimation. By satisfying these conditions, we can estimate the causal ARE reliably, focusing on obtaining the observable ARE efficiently without additional assumptions such as linearity. Although our identification assumptions may not always hold in all applications, we can at least eliminate model misspecification bias and minimize random variation. Once we have established how to estimate the ARE for a fixed region, we will turn our attention to the problem of finding a good region $\mathcal A$ and finally how to do that without incurring selection bias in estimating the ARE for that region.

\section{Estimating ARE with TMLE}

In the previous sections, we established that the causal ARE is equivalent to the observable ARE \(E[E[Y|T = 1, W] - E[Y]]\) under standard identification assumptions. Therefore, to estimate it, we need to: 1) create a new binary random variable \(T = 1_{\mathcal{A}}(A)\) and 2) proceed as if we were estimating the observable average counterfactual effect \(E[E[Y|T = 1, W]]\) from the observational data structure \((Y, \mathcal{A}, W)\) where \(\mathcal{A}\) is a simple binary "exposure".

We can adapt techniques for estimating the ATE \(E[E[Y|T = 1, W]] - E[E[Y|T = 0, W]]\) to construct ARE estimators that offer provable unbiasedness (assuming that there is no bias from potential violations of identification assumptions) and efficiency. These estimators are optimized for minimal sampling variance and possess the "doubly robust" property, as highlighted in \cite{Zheng2010, Zivich2021}. Both Augmented Inverse Propensity-Weighting (AIPW) and Targeted Maximum Likelihood Estimation (TMLE) stand out as assumption-free and data-efficient methods when used with machine learning and cross-fitting. Although they often produce similar results, TMLE has been found to perform more efficiently with smaller samples \cite{Luque-Fernandez2018, Smith2022, van2011targeted, van2018targeted}, making it a generally preferable approach.

A comprehensive analysis by Li et al. \cite{best_tmle} using ten different nutrition intervention studies showed that TMLE, especially in its cross-validated form (CV-TMLE), consistently provided robust and efficient estimates across various realistic simulation scenarios. The study concluded that the additional layer of cross-validation helps avoid unintentional over-fitting of nuisance parameter functionals, leading to more reliable inferences compared to other estimators. This reinforces the advantage of TMLE in practical applications, emphasizing its superior performance in maintaining efficiency and robustness in finite samples. Therefore, for estimating our data-adaptively identified region, we use the CV-TMLE approach.

The TMLE estimator is inspired by the fact that if we knew the true conditional mean \(Q(\mathcal{A}, W) = E[Y|T = 1, W]\), we could estimate the ARE with the empirical average \(\frac{1}{n}\sum_i Q(1, W_i) - Y_i\). Of course, we do not know \(Q\), but we can estimate it by regressing the outcome \(Y\) on the exposure \(\mathcal{A}\) and the covariates \(W\). However, using our estimate \(\hat{Q}\) instead of the truth incurs bias that might decrease as the sample size increases but dominates relative to random variability, making it impossible to establish p-values or confidence intervals. TMLE solves this problem by computing a correction to the regression model \(\hat{Q}\) that removes this bias, "targeting" the estimate \(\hat{Q}\) to the parameter of interest (here, ARE).

The process is as follows:
1. Use cross-validated ensembles of machine learning algorithms (a "super learner") to generate estimates of the conditional means of treatment: \(\hat{g}(\mathcal{A}=a, W) \approx P(\mathcal{A}=a | W)\) (i.e., propensity score) and outcome: \(\hat{Q}(\mathcal{A}, W) \approx E[Y | \mathcal{A}, W]\).
2. Regress \(Y\) (scaled to \([0,1]\)) onto the "clever covariate" \(H_i = \pi(\mathcal{A}_i) / \hat{g}(1, W_i)\) using a logistic regression with weights \(\text{logit}(\hat{Q}(\mathcal{A}, W))\). Here, \(\pi(\mathcal{A}_i)\) is the marginal probability of being in the region. The (rescaled) output of this is our target regression model \(\hat{Q}^*\).
3. Compute the plug-in estimate using the targeted model: \(\hat{\psi} = 1/n \sum_i \hat{Q}^*(1, W_i) - Y_i\).

An estimated standard error for \(\hat{\psi}\) is given by:
\[
\hat{\sigma}^2 = \frac{1}{n^2} \sum_i \left[\left(\frac{\pi(\mathcal{A}_i)}{\hat{g}(1, W_i)}\right)\left(Y_i - \hat{Q}^*(1, W_i)\right) + \left(\hat{Q}^*(1, W_i) - Y_i - \hat{\psi}\right)\right]^2.
\]

With the corresponding 95\% confidence interval \(\hat{\psi} \pm 1.96 \hat{\sigma}\).

Explaining why the targeting step takes the form of a logistic regression and how the estimated standard error is derived is beyond the scope of this work. \cite{vanderLaanRubin+2006, van2011targeted, van2018targeted} offer explanations targeted to audiences with varying levels of mathematical sophistication. To obtain these estimates, we need only to specify the ensemble of machine learning algorithms used to estimate the propensity and initial outcome regressions \(\hat{g}\) and \(\hat{Q}\). Theoretical guarantees hold as long as a sufficiently rich library is chosen. To estimate the ARE, we must also specify the region \(\mathcal{A}\) so that we can compute our binary "exposure" variable. This is the focus of the next section.

\section{Finding a Good Exposure Region}

Our target parameter $\psi$ depends on the region $\mathcal A$ to which we restrict the exposures (and, as such, should be denoted $\psi_{\mathcal{A}}$ to be precise). When such a region is not already provided (e.g. there are no existing thresholds for a multivariate exposure), it is natural to want to use the data to \textit{find} a region $A$ that would optimize population outcomes. In the case where the high value of $Y$ is considered harmful, the oracle parameter we seek can be formally described as

$$ \underset{\mathcal{A}}{\mathrm{argmin}} \int_w \int_a \mu(a,w) \frac{1_{\mathcal{A}}(a) p(a|w)}{\pi_{\mathcal{A}}(w)} p(w) \, da \, dw $$

or equivalently $\underset{\mathcal{A}}{\mathrm{argmin}} \ E[E[Y|T = 1, W]]$. 

When considering this oracle parameter, we also want to ensure the positivity assumption, specifically that the conditional probability of \( \mathcal{A} \) given \( W \) (i.e., \( P(\mathcal{A}|W) \)) is bounded away from 0 and 1, we want to avoid extreme propensity scores that could lead to instability or infinite variances in the estimates. This requirement reflects the need for sufficient variability for being inside and outside the exposure region across levels of covariates to make reliable causal inferences.

The oracle parameter in the given context is described as minimizing the expected value of the outcome \( Y \) conditional on being in an exposure region \( \mathcal{A} \) and given covariates \( W \). To incorporate the boundedness of \( P(\mathcal{A}|W) \), we adjust the formulation to explicitly ensure that the selected region \( \mathcal{A} \) does not result in probabilities of exposure that are too extreme. In this way, we can rewrite the oracle parameter, incorporating this boundedness directly into the optimization:

\[
\underset{\mathcal{A}}{\mathrm{argmin}} \int_w \int_a \mu(a,w) \frac{1_{\mathcal{A}}(a) p(a|w)}{\pi_{\mathcal{A}}(w)} p(w) \, da \, dw
\]
subject to the constraints:
\[
\epsilon \leq \pi_{\mathcal{A}}(w) \leq 1 - \epsilon, \quad \forall w
\]

Here, a reasonable $\epsilon$ may be 0.001, this ensures that no subgroup of  \( W \) is almost always or almost never exposed. This prevents the model from leveraging rare but extreme treatment effects that are not generalizable or might be based on spurious associations due to model overfitting or data anomalies.

Estimating this oracle region efficiently is extremely difficult. For example, even with access to the true regression function $ \mu$ and the conditional density function $ p(a|w)$ one would have to apply numerical integration (note $\pi_{\mathcal A}(w) = \int_{\mathcal A} p(a|w) da$) to compute plug-in estimates $\hat \psi_{\mathcal A}$ for some large number of candidate regions $\mathcal A$. It is not possible to efficiently search the set of all regions, even if each numerical integration was quick to compute. This problem is similar to challenges in scan statistics which often, rather than applying direct analytical solutions, employ some sort of search algorithm \cite{sim_annealing_scan, Tango2005, Kulldorff1997}.

Thus, some restriction of the admissible regions is required, e.g. restricting $\mathcal A$ to be axis-aligned rectangles in the exposure space (this restriction also benefits the interpretability of the oracle region). Additional restrictions are also required to make the problem sensible when there is a point minimum in the conditional mean $\mu$. For example, consider a unidimensional exposure $a\in \mathbb R^1$ and let $\mu(a,w) = a^2$. In this case, the minimizing region is the single point $\mathcal A = \{0\}$. This single-point case is covered by our positivity restriction to the oracle parameter. Given restrictions on the region, we create a greedy approximation of the oracle region using recursive partitioning of the exposure space, wherein a partition is only made if the conditional probability of being exposed to a region is bounded away from 0 and 1, this algorithm is described next. 

\subsection{Targeted Decision Trees}

We introduce a novel decision tree algorithm developed to approximate the oracle parameter. This parameter, denoted as $\underset{\mathcal{A}}{\mathrm{argmin}} \ E[E[Y|T = 1, W]]$, represents the region $\mathcal{A}$ that minimizes the expected outcome for a given set of covariates $W$. Our approach is tailored to meet several key criteria: computational efficiency, interpretability, and implicit regularization to avoid overly granular and statistically insignificant regions. The core of our method is a regression tree that optimizes a simple plug-in estimator for the policy effect ($E[Y| T = 1, W]$) of a given axis-aligned rectangular region, based on a Random Forest or a similar straightforward model paired with targeted learning. The decision tree is constructed to search for the minimum/maximum region using this plug-in estimator, greedily evaluating splits in the exposure variables $A$ and selecting the region that maximizes/minimizes the estimated average outcome in the newly formed regions. 

Simply using g-computation to estimate the conditional mean outcome given a region and covariates may run into issues due to positivity. In areas of the exposure space that are very unlikely for certain $w$, our estimate of the mean outcome may behave in strange ways, especially when using nonparametric estimators. Model extrapolation could lead to very wild estimates in certain regions, but the data adaptive region might be selected  specifically because they have large (or small) estimates of the mean outcome. To embed in our algorithm this restriction of only selecting regions with support, we incorporate two restrictions 1. Partitioning occurs only if the conditional probability of being in the region is bounded away from 0 and 1 and 2. we debias our initial plug-in estimates of the conditional mean in the left and right regions of the tree using targeted learning and determine if there is a significant difference compared to the parent node mean by calculating the influence function for the population intervention effect, comparing the region means to the parent mean and getting estimates of variance for p-value calculations using the influence function. We give a detailed pseudocode of the algorithm in \textbf{Algorithm 1}, a brief synopsis is given below. 

In the partitioning criterion phase, the algorithm explores all binary splits across the range of each exposure variable to identify the optimal partitioning strategy. Consider a scenario where we have five exposure variables, A1 through A5. The algorithm examines the unique values for each exposure variable. For example, for A1, it evaluates the potential splits at A1 < 1, A1 < 2, and so forth, up to A1 < m, where m represents the maximum observed value of A1. At each of these split points, the algorithm constructs a binary indicator defining the region of interest and employs the plug-in estimator. This estimator is trained to assess the mean outcome under the hypothetical situation where all individuals are `forced' into that region. The regions are to the left and right of the split point. For each, we debias the initial estimate using targeted learning and assess the significance of the region mean relative to the parent node mean by estimating the influence function for the population intervention effect, comparing the region node means to the parent node mean.

The algorithm's objective at each split point is to minimize/maximize the average predicted outcome within the newly formed child nodes, relative to the parent node. This process of finding the optimal split is applied to each of the exposure variables. For example, if the algorithm identifies that splitting at A5 < 5 yields the minimum average outcome, this split becomes the root node of our decision tree. Subsequently, the algorithm partitions the data into two segments based on the chosen split, resulting in a left node (A5 $<$ 5) and a right node (A5 $\geq$ 5). The recursive process is then applied to each of these data regions demarcated by the node, iterating through all exposure variables again to find further splits that minimize the average outcome. For example, within the left node, the algorithm might test combinations such as A5 $<$ 5 $\&$ A1 $<$ 2, etc., each time training the plug-in estimator and evaluating the mean outcome for the defined region. This recursive partitioning continues until two stopping criteria are met: it either reaches the pre-specified maximum depth of the tree, or it arrives at a point where no further split can yield an average outcome significantly lower/greater than that of the parent node. 

\begin{algorithm}
\caption{Recursive Partitioning Decision Tree using TMLE}
\begin{algorithmic}[1]
\State \textbf{Input:} Data $D$, exposures $\mathbf{A}$, covariates $W$, outcome $Y$, max depth $d_{\text{max}}$, min or max
\State \textbf{Output:} Rule that demarcates the region optimizing $E[Y | \mathcal{A}^* = 1, \boldsymbol{A}, W]$ 

\Procedure{TargetPart}{$D$, $\mathbf{A}$, $W$, $Y$, depth=0}
\If{depth $= d_{\text{max}}$}
    \State \textbf{return} $best\_split$ \Comment{Stop recursion}
\EndIf
\State Init $best\_split \gets \text{Null}$, $p_{\text{best}} \gets 1$, $min\_avg \gets \infty$, $max\_avg \gets -\infty$

\For{$a \in \mathbf{A}$}
    \For{$s$ in unique values of $a$}
        \State $\mathcal{A}^* \gets (D[a] \leq s)$
        \State Fit $P(\mathcal{A}^*|W)$; get propensities $p$
        \If{any p < 0.001 \textbf{or} $p$ > 0.99}
  \Comment{continue}
\EndIf
        \State Calculate $H(a,w)$ from $p$
        \State Fit outcome model $E[Y| \mathcal{A}^*, \boldsymbol{A}, W]$; predict for $\mathcal{A}^* = 1, 0$
        \State Update predictions (TMLE), compare left and right region averages to parent average, $\theta_{\text{left}}, \theta_{\text{right}}$
        \State Calculate SEs from IC, p-values; update $best\_split$ if p-value $<$ $p_{\text{best}}$    \EndFor
\EndFor

\If{$best\_split$}
    \State $D_{\text{left}} \gets D[\mathcal{A}^* = 1]$, $D_{\text{right}} \gets D[\mathcal{A}^* = 0]$
    \State Recursively apply \Call{TargetPart}{data subset} based on $best\_split$
\EndIf
\EndProcedure

\State Initiate \Call{TargetPart}{$D$, $\mathbf{A}$, $W$, $Y$, 0}
\end{algorithmic}
\end{algorithm}

\section{K-fold Cross-Estimation}

As discussed, the identification of interaction regions and the estimation of ARE requires a data partitioning strategy that mitigates bias. We employ a K-fold cross-estimation technique that ensures the asymptotic properties of our estimators without reliance on additional assumptions. This involves dividing the data set into complementary estimation ($P_{n_k}$) and parameter-generating ($P_{n_{-k}}$) samples, the latter being utilized to derive exposure regions and train nuisance parameters essential for the target maximum likelihood estimate (TMLE) update. For each fold, we determine the exposure thresholds using $P_{n_{-k}}$ and, with the same sample, train our nuisance estimators $g_n$ and $Q_n$. Subsequently, these estimators are applied to $P_{n_k}$ to obtain unbiased ARE estimates within the fold. 

\subsection{Pooled TMLE}

Upon completion of the cross-estimation procedure, a pooled TMLE update provides a summary measure of the oracle region target parameter across k-folds. Specifically, we stack the predictions for each nuisance parameter from the estimation samples and run a pooled TMLE update on the cumulative initial estimates. The resulting average is then used for parameter estimation. 

Our substitution estimator, denoted by $\bar{Q}_n$, approximates the true conditional mean $\bar{Q}_0$ by plugging in the empirical distribution for each observation. The estimator is operationalized via a Super Learner algorithm followed by TMLE, and its cross-validated variant is expressed as 

$$\Psi(Q_{P_{n_k}}^{\star}) = \frac{1}{V} \sum_{v=1}^V \{ \bar{Q}_{n_{-k}}^{\star}(T_{n_{-k}} = 1, W_v) - Y_v\}$$. 

Where $Q_{n_{-k}}^{\star}$ is the TMLE updated expectation of the outcome if everyone were exposed to levels of exposure within the data-adaptively determined region. The cross-estimation approach not only utilizes the full dataset for variance estimation, resulting in tighter confidence intervals, but also enables the derivation of fold-specific and pooled estimates. 

That is, this pooled parameter represents the ARE for the determined maxmizing/minimizing region in the exposure space. Of course, the actual regions determined in folds can vary, and thus this acts as an omnibus test, which can be used to evaluate fold-specific results. A significant pooled TMLE ARE for the oracle region can be interpreted as ``there is a region in the exposure space where the expected outcome is significantly different compared to the average outcome''. The analyst can then evaluate the fold-specific results to determine what regions make up this oracle estimate. Of course, interpretability of the pooled oracle estimate is reliant on regions in the same sets of exposures that are found across the folds. Otherwise, this pooled estimate allows the analyst to understand if there is generally "signal" in the data for a region or set of regions that collectively differ in expected outcomes compared to the average. This dual presentation allows for the evaluation of variability across folds, offering insight into the stability of the pooled ARE. In particular, when exposure regions $\mathcal{A}$ exhibit high variability between folds, fold-specific results provide an essential interpretive counterbalance to the pooled estimates. 

\subsection{Theoretical Properties of Region Holding and ARE Convergence}

Theoretical results for data-adaptive parameters are shown in \cite{Hubbard2016} which established the asymptotic properties of these parameters under a framework that leverages cross-validation through data partitioning into \(V\) equally sized sub-samples.

The estimator developed for each partition is not only consistent but also asymptotically normal due to the independence of the partitioned samples and the efficient estimator applied within each partition. They demonstrate that the average of these estimators, denoted as \(\hat{\psi}_{n}\), converges in distribution to a normal distribution centered at the true data-adaptive parameter \(\psi_{0}\), with a variance that can be consistently estimated. This convergence is underpinned by the central limit theorem, which applies due to the independent and identically distributed nature of the estimators across partitions.

Furthermore, it is crucial to address the role of the restriction on the size of the exposure region in our case, based on the positivity assumption. This assumption ensures that each exposure level within the region has a non-negligible probability of occurrence, which is critical for maintaining the validity of statistical inferences as the sample size increases. Without this assumption, as sample sizes grow, the selected region might progressively shrink, potentially invalidating the model due to the violation of the positivity condition. Hence, maintaining a sufficiently large region is vital for ensuring that the estimators converge appropriately and that the asymptotic properties hold, thereby avoiding biases that could arise from increasingly narrow exposure definitions.

\section{Simulations}
In this section, we demonstrate using simulations that our approach identifies the correct exposure region that maximizes/minimizes conditional mean and estimates the correct PIE target parameter built into a DGP for this region. Henceforth we refer to our method as CVtreeMLE.

\subsection{Data-Generating Processes}

Because a two-dimensional exposure space is easier to visualize and describe compared to higher-dimensional spaces, we created a squared dose-response relationship between two exposure variables where an interaction occurs between the exposures when each meets a particular threshold value. Specific outcome values were generated for each subspace of the mixture based on split points $\mathcal{A}_d$, but there exists a region with the maximum outcome (the truth that we want). The goal is to determine whether our adaptive target parameter is targeting the region that maximizes/minimizes the conditional mean outcome for the given sample and to evaluate how \texttt{ CVtreeMLE} approaches this desired oracle parameter (the true PIE for this region) as the sample size increases. To achieve this goal, we construct a data-generating process (DGP) explained below.

\subsubsection{Two-Dimensional Exposure Simulations}

This DGP has the following characteristics, $O = (W,A,Y)$. $W$ are three baseline covariates, $W_1 \sim \mathcal{N}(\mu = 37, \sigma = 3), W_2 \sim \mathcal{N}(\mu = 20, \sigma = 1), W_3 \sim \mathcal{B}(\mu = 0.5)$. Here $ \mathcal{B}$ is a Bernoulli distribution, and $\mathcal{N}$ is normal. These distributions and values were chosen to represent a study with covariates for age, BMI, and sex. Our generated exposures were also created to represent two chemical exposures quantized into five discrete levels. The values and range of the outcomes were chosen to represent common environmental health outcomes, such as telomere length or epigenetic expression. We are interested in sampling observations into a 2-dimensional exposure grid. We use discrete exposures to facilitate calculation of the ground truth effects without numerical integration. Here, a $5 \times 5$ grid is based on combinations of two discrete exposure levels with values 1-5. We want the number of observations in each of these cells to be affected by covariates. To do this, we define a conditional categorical distribution $P\{(A_1, A_2)=(a_1, a_2) | W=w\}$ and sample from it. $$P\{(A_1, A_2) = (a_k, a_l) | W\} = \frac{e^{W^\top \beta_{k,l}}}{1 + \sum_{k,l} e^{W^\top \beta_{k,l}}}$$ Here, the $\beta$'s attached to each covariate were drawn from a normal distribution with means 0.3, 0.4, 0.5 and 0.5, respectively, all with a standard deviation of 2.  This then gives us 25 unique exposure regions with densities dependent on the covariates. We then want to assign an outcome to each of these regions based on the main effects and interactions between exposures. We generate the outcome as:  $Y = 0.2 A_1^2 + 0.5 A_1 A_2 + 0.5 A_2^2 + 0.2 * W_1 + 0.4 * W_3 + \mathcal N(0, 0.1)$. This indicates that there is a slightly weaker squared effect for $A_2$ relative to $A_1$ and a strong interaction between the exposures and confounding due to $W_1$ and $W_2$.

\paragraph{Computing Ground Truth}
The fact that our exposures are discrete in this simulation allows us to easily compute the ground truth mean outcome for any region $\mathcal A$ because we can explicitly compute the conditional mean function $Q_\mathcal{A}$
\begin{align*}
Q(\mathcal A=1, w)
&=
\int_{a \in \mathcal A}
 \mu(a,w) 
 \frac
{p_{A|W}(a,w)}
{\pi_{\mathcal A}(w)} \ da
\\
 &=
 \frac
{\sum_{a \in \mathcal A} \mu(a,w) p_{A|W}(a,w)}
{\sum_{a\in \mathcal{A}} p_{A|W}(a)}
\\
\end{align*}

and select the region that matches our oracle parameter under our set of positivity restrictions. Therefore, to approximate the ARE of a region $\mathcal{A}$ to arbitrary precision we compute $$\text{ARE}(\mathcal A) = \frac{1}{b}\sum_i^b Q_\mathcal{A}(1,W_i) - Y_i$$ using a large $b$. We repeat this for each of the $(6^2)^2$ possible rectangular regions to determine which has the optimal ARE.

\subsection{Evaluating Performance}

The following steps describe how our simulation was tested to determine 1. asymptotic convergence to the true oracle region used in the DGP, 2. convergence to the true ARE based on this oracle region and 3. convergence to the true data-adaptive ARE, that is, \texttt{ CVtreeMLE}'s ability to correctly estimate the ARE if the data-determined rule was applied to the population. We do this by first generating a random sample from the DGP of size $n$ which is broken into $K$ equal size estimation samples of size $n_k = n/K$ with the corresponding parameter that generates samples of size $n-n/K$. At each iteration, the parameter generating fold is used to define the region via our algorithm and is used to estimate the necessary nuisance estimators $\hat Q$ and $\hat g$. The estimation fold is used to get our TMLE updated estimate, and then we do this for all folds. For an iteration, we output the ARE estimates given the pooled TMLE and the k-fold specific TMLE. We compare the estimated and ground truth AREs for each region identified in each fold to estimate the data-adaptive bias. We also compare estimated AREs and regions with oracle equivalents.

For each iteration, we calculate metrics for bias, variance, MSE, CI coverage, and confusion table metrics for the true minimal region compared to the estimated region. We also calculate the bias for the estimated mean in the region found in our recursive partitioning algorithm to find the minimum region. For each type of estimate (pooled TMLE, k-fold specific TMLE estimates), we have bias when comparing our estimate to 1. The ARE based on the true region in the DGP that minimizes the outcome mean 2. The ARE when the data-adaptively determined region is applied to the population. Therefore, when comparing to the true "oracle" region ARE we have $\psi^0_{\text{pooled tmle MSE}}$: This is the MSE of the pooled TMLE ARE compared to the ground-truth ARE for the true oracle region built into the DGP which minimizes the mean adjusted outcome and $\psi^0_{\text{mean v-fold tmle MSE}}$: This is the MSE of the mean k-fold specific AREs compared to the ground-truth ARE for the true region built into the DGP.

We are also interested in the ARE if the data-adaptive-determined region, the region estimated to minimize the adjusted outcome in the sample data, were applied to $P_0$ the true population. Therefore, there are also MSE estimates for $\psi^{DA}_{\text{pooled tmle MSE}}$: This is the MSE of the pooled TMLE ARE compared to the ARE of the average region across the folds applied to $P_0$ and $\psi^{DA}_{\text{mean v-fold tmle MSE}}$: This is the MSE of the mean k-fold specific AREs compared to the mean ARE when all the k-fold specific rules are applied to $P_0$. 

For each ARE estimate we calculate the confidence interval coverage of the true ARE parameter given the oracle region and the ARE given the data-adaptively determined region applied to $P_0$. For the TMLE pooled estimates, these are lower and upper confidence intervals based on the pooled influence curve. For the k-fold-specific coverage, we take the mean lower and upper bounds.  In each case, we check to see if the ground-truth rule ARE and the data-adaptive rule ARE are within the interval.

Lastly, we compare the data-adaptively identified region (the region determined by our recursive partitioning algorithm to be the minimal region amongst all regions) with the ground-truth region (built into the DGP) using the confusion table metrics for true positive, true negative, false positive, and false negative to determine whether, as the sample size increases, we converge to the true region. Similarly, we compare the expected outcome for the minimum region calculated in our recursive partitioning algorithm to the true minimum outcome to ensure that our estimates are converging to the truth. These performance metrics were calculated in each iteration, where 100 iterations were performed for each sample size $n =$ (200, 350, 500, 750, 1000, 1500, 2000, 3000, 5000). It was ensured that for each data sample, at least one observation existed in the ground truth region to ensure that confusion table estimates could be calculated. CVtreeMLE was run with a 10-fold CV with default learner stacks for each nuisance parameter and data-adaptive parameter. 

\subsection{Default Estimators}
As discussed, \texttt{CVtreeMLE} needs estimators for $Q  = E(Y|A,W)$ and $g = P(A|W)$. \texttt{CVtreeMLE}  has built in default algorithms to be used in a Super Learner \cite{SL_2008} that are fast and flexible. These include random forest, general linear models, elastic net, and xgboost. These are used to create Super Learners for $Q_n$ and $g_n$.  Users can pass in their own libraries for these nuisance and data-adaptive parameters. For these simulations, we use these default estimators in each Super Learner. Our partitioning algorithm to discover the region $\mathcal A$ was set to have a maximum depth of 3 and a minimum number of observations per leaf to 10. In this simulation we are interested in finding the region with maximum outcome. The package has a "min\_max" parameter which we set to "max" to find the region with maximum expected mean outcome.

\subsection{Results}

\subsubsection{CVtreeMLE Algorithm Identifies the Oracle Region with Minimum Expected Outcome}

\begin{figure}[!h]
  \hspace*{2 cm}\includegraphics[scale = 0.4]{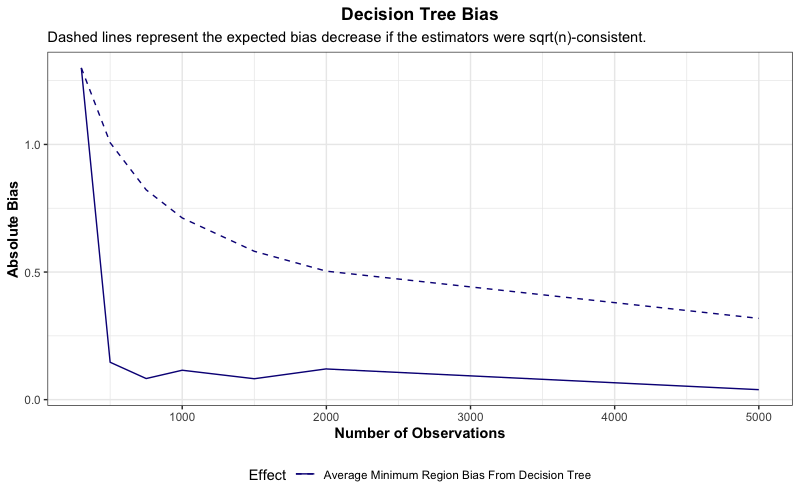}
    \vspace{-1cm}
  \caption{Average Bias from G-computation Plug-in from the Decision Tree Algorithm}
  \label{fig:decision_tree_bias}
\end{figure}

First, we describe the results to identify the true region built into the DGP. It is obviously necessary that this converges to the truth as the sample size increases in order for the estimates $\psi^0$ to be asymptotically unbiased. Overall, we find that our tree algorithm identifies the true region in the DGP and therefore provides results that have a minimum outcome for treatment policies. Our recursive algorithm shows very good identification of the true outcome minimizing region even in low sample sizes. At sample size 300 the average false positives were 0.053, at 500, 0.0005, and at 750 onward 0. The average true positives were 100\% across all sample sizes. Our recursive partitioning algorithm correctly identifies the interaction between both exposures as the minimizing region (as opposed to marginal impacts) in 100\% of folds in all sample sizes.  \textbf{Figure \ref{fig:decision_tree_bias}} shows the bias in comparing the estimated average in the estimated minimum region of the decision tree to that of the true oracle average in the oracle minimizing region. From this figure, it can be seen that we have asymptotic convergence to the true minimal average at a rate faster than $\sqrt{n}$ for this simulation.

\subsubsection{CVtreeMLE Unbiasedly Estimates the Data-Adaptive Parameter}
\begin{figure}[!h]
  \hspace*{-2.5cm}\includegraphics[scale = 0.33]{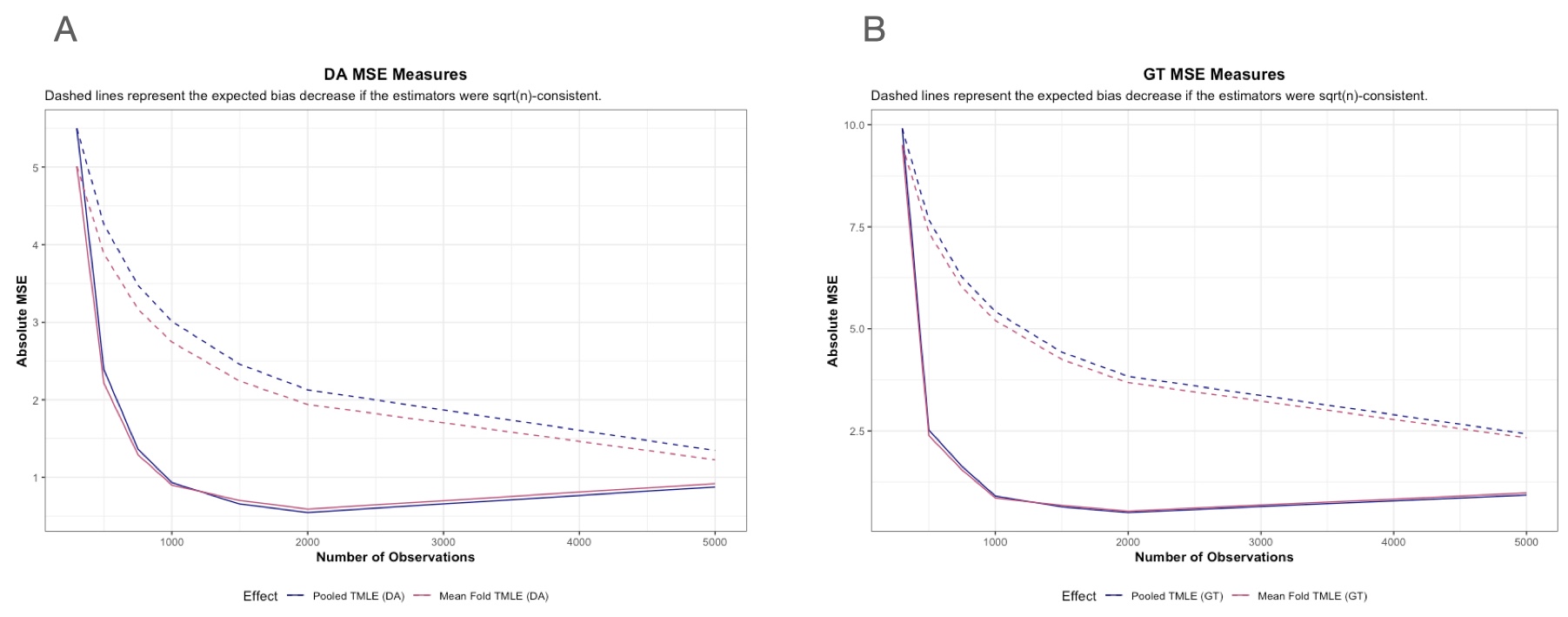}
  \caption{Data-Adaptive and Oracle Parameter MSE}
  \label{fig:mse_compiled}
\end{figure}

Looking at the MSE for the ARE estimate between the adaptive region discovered in the data and the ARE when this region is applied to $P_0$, TMLE unbiasedly estimates the adaptive parameter of the data at root rates $n$ with an average of 95\% CI coverage. Below, \textbf{Figure \ref{fig:mse_compiled} A} shows the data-adaptive rule ARE MSE ($\psi^{DA}_{\text{pooled tmle MSE}}$, $\psi^{DA}_{\text{mean k-fold tmle MSE}}$). Given that the pooled TMLE is the average estimates across folds after the pooled update, we would expect these estimates to be fairly close, which they are. Coverage by sample size was: 300 = 98\%, 750 = 94\%, 2000 = 95\%, 5000 = 95 \%. 

\subsubsection{CVtreeMLE Unbiasedly Estimates the Oracle Target Parameter}

Now, we compare the estimates with the true ARE given the oracle region in the DGP.  \textbf{Figure \ref{fig:mse_compiled} B} show the MSE for this comparison in our simulation. As can be seen, both decrease faster than the root $n$ rate for both the fold-specific and pooled estimates. Based on these simulations, \texttt{CVtreeMLE} unbiasedly estimates the oracle target parameter at root $n$ rates.  Coverage by sample size was: 300 = 97\%, 750 = 95\%, 2000 = 94\%, 5000 = 95 \%.

\section{Applications}

\subsection{NIEHS Synthetic Mixtures}

The NIEHS synthetic mixture data (found here in \href{https://github.com/niehs-prime/2015-NIEHS-MIxtures-Workshop}{github}) are a commonly used data set to evaluate the performance of statistical methods for mixtures. These synthetic data can be considered as the results of a prospective cohort study. The outcome cannot cause exposures (as might occur in a cross-sectional study). Correlations between exposure variables can be thought of as caused by common sources or modes of exposure. The nuisance variable $W$ can be assumed to be a potential confounder and not a collider. There are 7 exposures ($A_1 - A_7$) that have a complicated dependency structure with a biologically based dose response function based on endocrine disruption. For details, the synthetic data key on the github page for data set 1 (used here) gives a description of how the data were generated. Generally, there are two groups of exposures ($A_1, A_2, A_3$ and $A_5, A_6$). Therefore, the correlations within these clusters are high. $A_1, A_2, A_7$ contribute positively to the outcome; $A_4, A_5$ contribute negatively; $A_3$ and $A_6$ do not have an impact on the outcome, making rejecting these variables difficult given their correlations with members of the group. This correlation and effects structure is biologically plausible because different congeners of a group of compounds (e.g., PCBs) may be highly correlated but have different biological effects. There are various agonistic and antagonistic interactions in exposures. \textbf{Table \ref{tab:NIEHS_intxns}} shows the interactions built into this synthetic data. 

\begin{table}[h]
\centering
\begin{tabularx}{\textwidth}{l|X} 
\toprule
Variables & Interaction Type                                                                          \\ 
\midrule
A1 and A2 & Toxic equivalency factor, a special case of concentration addition (both increase Y) \\
A1 and A4 & Competitive antagonism (similarly for A2 and A4)                                          \\
A1 and A5 & Competitive antagonism (similarly for A2 and A4)                                          \\
A1 and A7 & Supra-additive (“synergy”) (similarly for A2 and A7)                                      \\
A4 and A5 & Toxic equivalency factor, a type of concentration addition (both decrease y)              \\
A4 and A7 & Antagonism (unusual kind) (similarly for A5 and A7) \\ 
\bottomrule                                     
\end{tabularx}
\caption{NIEHS Synthetic Data Interactions}
\label{tab:NIEHS_intxns}
\end{table}

Given these toxicological interactions, we can expect certain statistical interactions determined as cut-points for sets of variables from CVtreeMLE. Since we are interested in one region that minimizes the mean outcome under some restrictions, we expect to find a region that includes lower levels for an exposure with an agonistic relationship or perhaps an interaction between an agonistic and antagonistic exposure. For example, the synergistic relationship between A1 and A7 may indicate we should decrease these to a certain level simultaneously. The NIEHS data set has 500 observations and 9 variables. $A$ is a binary confounder. We apply CVtreeMLE to this NIEHS synthetic data using a 10-fold CV and the default stacks of estimators used in the Super Learner for all parameters. We parallelize over the cross-validation to test computational run-time on a newer personal machine an analyst might be using. We show results when searching for the minimizing region.

\subsection{NIEHS Data Results}
In 10 of 10 folds, it was found that a region in $X_1 \& X7$ had the lowest adjusted outcome. \textbf{Table \ref{tab:niehs_k_are_effects}} shows the k-fold specific results. In this table, the ARE is the change in outcome if all individuals were in the region defined in the column ''region'' compared to the observed outcome. Standard errors and confidence intervals are derived from the fold-specific influence function based on the validation data. 

\begin{table}[ht]

\caption{NIEHS K-fold ARE Results}
\hspace{-1cm}\begin{tabular}{crrrrrrl}
\hline
\textbf{Fold} & \textbf{ARE} & \textbf{SE} & \textbf{Lower CI} & \textbf{Upper CI} & \textbf{P-value} & \textbf{Adjusted P-value} & \textbf{Region} \\ 
\hline
1 & -4.33 & 27.86 & -58.94 & 50.29 & 0.88 & 1.00 & $X1 \leq 0.98 \& X7 \leq 0.5$ \\
2 & -2.88 & 16.90 & -36.00 & 30.25 & 0.86 & 1.00 & $X1 \leq 1 \& X7 \leq 0.45$ \\
3 & -2.39 & 20.04 & -41.66 & 36.88 & 0.91 & 1.00 & $X1 \leq 1 \& X7 \leq 0.46$ \\
4 & -3.49 & 24.92 & -52.34 & 45.35 & 0.89 & 1.00 & $X1 \leq 1 \& X7 \leq 0.33$ \\
5 & -3.88 & 23.85 & -50.63 & 42.87 & 0.87 & 1.00 & $X1 \leq 1 \& X7 \leq 0.44$ \\
6 & -2.98 & 21.85 & -45.81 & 39.85 & 0.89 & 1.00 & $X1 \leq 1 \& X7 \leq 0.52$ \\
7 & -4.12 & 18.73 & -40.83 & 32.58 & 0.83 & 1.00 & $X1 \leq 0.99 \& X7 \leq 0.35$ \\
8 & -4.52 & 25.85 & -55.18 & 46.14 & 0.86 & 1.00 & $X1 \leq 1 \& X7 \leq 0.4$ \\
9 & -3.98 & 20.39 & -43.95 & 35.98 & 0.85 & 1.00 & $X1 \leq 1 \& X7 \leq 0.42$ \\
10 & -3.13 & 27.96 & -57.93 & 51.68 & 0.91 & 1.00 & $X1 \leq 1 \& X7 \leq 0.4$ \\
\hline
\end{tabular}
\label{tab:niehs_k_are_effects}
\end{table}

Because these K-fold estimates are for very similar regions in $X1$ and $X7$ the pooled TMLE parameter is interpretable. \textbf{Table \ref{tab:niehs_mixture_ate}} shows the pooled results. Here, the ``Mixture ARE'' column is the average ARE estimates across the folds. The standard error and confidence intervals are now derived from all the validation fold estimates. 

\begin{table}[ht]
\caption{Summary of mixture average treatment effects (ATE) across different folds.}
\hspace{-2cm}\begin{tabular}{crrrrrp{4cm}c}
\hline
\textbf{Fold} & \textbf{Mixture ARE} & \textbf{Standard Error} & \textbf{Lower CI} & \textbf{Upper CI} & \textbf{P-value} & \textbf{Average Rule} \\ 
\hline
Pooled TMLE & -3.73 & 7.31 & -18.06 & 10.59 & 0.61 & $X1 \leq 0.997$ $(0.98,1)$ \& $X7 \leq 0.427$ $(0.33,0.52)$ \\
\hline
\end{tabular}
\label{tab:niehs_mixture_ate}
\end{table}

The average rule was found to be $X_1 <= 0.997 \& X_7 <= 0.43$. This threshold varied between 0.98 and 1  for $X_1$ and 0.33 and 0.52 for $X_7$, a very low partition variability. The ARE for this region was found to be -3.73 (--18.06 - 10.59), which means that if all individuals were exposed to levels of $X_1$ and $X_7$ within this region, the endocrine disrupting outcome would be 3.73 units less than the observed average. This effect was not significant. The fact that this region was found in 100\% of folds indicates a robust result. This result can be interpreted as, in order to have the largest impact to reduce the endocrine disrupting outcome, this exposure should be adjusted to levels below this threshold. According to the description of this data found here on \href{https://github.com/niehs-prime/2015-NIEHS-MIxtures-Workshop}{github} the exposures $X_1$ and $X_7$ have the strongest synergistic interaction. Therefore, our findings make sense insofar as, to have the greatest reduction in the outcome, both exposures must be reduced simultaneously. 

\subsubsection{Comparison to Existing Methods}
Quantile g-computation \cite{Keil2019}, prevalent in environmental epidemiology for mixture analysis, estimates the effects of uniformly increasing exposures by one quantile, based on linear model assumptions. This method quantizes mixture components, summing the linear model's coefficients to form a summary measure ($\Psi$) for joint impact assessment. However, it inherently assumes additive, monotonic exposure-response relationships, overlooking complex, potentially nonlinear interactions typical in mixtures, like in endocrine disrupting compounds. Consequently, this method might not accurately capture the nuanced dynamics of mixed exposures, especially when interactions vary with other variable levels. 

We run quantile g-computation on the NIEHS data using 4 quantiles with no interactions to investigate results using this model. The size of the scaled effect (positive direction, sum of positive coefficients) was 6.28 and included $A_1, A_2, A_3, A_7$ and the scaled effect size (negative direction, sum of negative coefficients) was -3.68 and included $A_4, A_5, A_6$. Compared to NIEHS ground truth, $A_3, A_6$ are incorrectly included in these estimates. However, the positive and negative associations for the other variables are correct. Next, because we expect interactions to exist in the mixture data, we would like to assess for them but the question is which interaction terms to include? Our best guess is to include interaction terms for all exposures. We do this and show results in \textbf{Table \ref{tab:q_comp_itxns}}.

\begin{table}[ht]
\centering
\begin{tabular}{rrrrrr}
  \toprule
 & Estimate & Std. Error & Lower CI & Upper CI & Pr($>$$|$t$|$) \\ 
  \midrule
(Intercept) & 21.29 & 1.58 & 18.19 & 24.39 & 0.00 \\ 
  psi1 & 0.02 & 1.62 & -3.16 & 3.20 & 0.99 \\ 
  psi2 & 0.59 & 0.67 & -0.71 & 1.90 & 0.37 \\ 
   \bottomrule
\end{tabular}
\caption{Quantile G-Computation Interaction Results from NIEHS Synthetic Data}
\label{tab:q_comp_itxns}
\end{table}

In \textbf{Table \ref{tab:q_comp_itxns}} $\Psi_1$ is the summary measure for the main effects and $\Psi_2$ for interactions. As can be seen, when including all interactions, neither of the estimates are significant. Of course, this is to be expected given the number of parameters in the model and the sample size $n=500$. However, moving forward with interaction assessment is difficult; if we were to assess for all 2-way interaction of 7 exposures, the number of sets is 21 and with 3-way interactions is 35. We would have to run these many models and then correct for multiple testing. Hopefully, this example shows why mixtures are inherently a data-adaptive problem and why popular methods such as this, although succinct and interpretable, fall short even in a simple synthetic data set. 

\subsection{NHANES Data}
Chemical and metal exposures are linked to changes in telomere length, a key factor in cellular aging and disease. Our applied study explores the collective influence of metal exposure on LTL, aiming to: 1) Present CVtreeMLE application results on real-world data, 2) Share data and processing code via the open-source CVtreeMLE package, and 3) Identify the most harmful exposure levels on telomere length. We utilize data from the NHANES 1999-2002, encompassing demographic information, disease history, nine urinary metals, and LTL for 2510 participants. Confounders include age, gender, race, and lifestyle factors. Metals measured are barium, cadmium, cobalt, cesium, molybdenum, lead, antimony, thallium, and tungsten. Our analysis with CVtreeMLE involves a 10-fold CV and seeks regions where metal exposure most reduces LTL. Data and more information are available in the CVtreeMLE package for further research.

\begin{table}[ht]
\caption{Mixtures of Metals and Telomere Length Across Different Folds}
\label{tab:metals-telomere}
\hspace{-1cm}\begin{tabular}{ccccccp{9cm}}
  \hline
  Fold & ARE & SE & Lower CI & Upper CI & p-value & Mix Rule \\ 
  \hline
  1 & -0.54 & 0.02 & -0.58 & -0.50 & p < 0.001 & barium > 0.37 \& molybdenum $\leq$ 112.4 \\
  2 & -0.68 & 0.02 & -0.72 & -0.64 & p < 0.001 & barium > 0.39 \& molybdenum $\leq$ 110 \\
  3 & -0.99 & 0.02 & -1.03 & -0.95 & p < 0.001 & barium > 0.42 \& molybdenum $\leq$ 113.9 \\
  4 & -0.26 & 0.02 & -0.29 & -0.22 & p < 0.001 & barium > 0.41 \& barium > 0.63 \\
  5 & -0.60 & 0.02 & -0.64 & -0.56 & p < 0.001 & barium > 0.41 \& molybdenum $\leq$ 109.7 \\
  6 & -0.69 & 0.02 & -0.74 & -0.65 & p < 0.001 & barium > 0.43 \& molybdenum $\leq$ 113.3 \\
  7 & -2.58 & 0.05 & -2.68 & -2.48 & p < 0.001 & barium > 0.4 \& cesium > 2.23 \\
  8 & -0.96 & 0.02 & -1.00 & -0.92 & p < 0.001 & barium > 0.4 \& cesium > 2.1 \\
  9 & -0.70 & 0.02 & -0.75 & -0.66 & p < 0.001 & barium > 0.42 \& molybdenum $\leq$ 119.2 \\
  10 & -0.71 & 0.02 & -0.75 & -0.67 & p < 0.001 & barium > 0.42 \& molybdenum $\leq$ 112 \\
  \hline
\end{tabular}
\end{table}

\subsection{NHANES Mixed Metal Results}
\textbf{Table \ref{tab:metals-telomere}} shows the k-fold specific results.
The results reveal significant associations in each fold with consistent patterns of metal mixtures leading to decreased telomere length. For example, higher levels of barium combined with varying thresholds for lower molybdenum were consistently associated with reduced telomere lengths, as evidenced by consistently negative AREs across several folds. In particular, the combination of barium (>0.42) and molybdenum (<113.9) in Fold 3 showed one of the most substantial decreases in telomere length (ARE = -0.99), with all reported p-values less than 0.001, which confirms the robustness of these findings. This pattern suggests a repeatable dose-response relationship in which specific levels of barium and molybdenum contribute to the shortening of the telomere. In one fold, fold 7, the minimizing region was in barium and cesium and in one fold, fold 4, the minimizing region was for barium alone without interaction. The fact that the minimizing region was found for high levels of barium (>0.4) and low levels of molybdenum < 113 in 7 of 10 folds indicates that these two chemicals have an antagonistic relationship which in combination can lead to the most severe reduction in telomere length. Running in more folds (such as 20) could make this result more consistent. 

\begin{table}[ht]
\centering
\caption{Pooled Estimation of Regional Attributable Effect}
\label{tab:regional-are}
\begin{tabular}{cccccc}
  \hline
  \textbf{Fold} & \textbf{Region ARE} & \textbf{Standard Error} & \textbf{Lower CI} & \textbf{Upper CI} & \textbf{P-value} \\ 
  \hline
  Pooled TMLE & -2.52 & 0.01 & -2.54 & -2.50 & $<$0.001 \\ 
  \hline
\end{tabular}
\end{table}

The ARE for the oracle parameter across the folds was found to be -2.52 (-2.54 - -2.50), p < 0.001. This indicates that for a region in the metal mixture space that minimizes telomere length, if all individuals were exposed to this region the telomere length would reduce by 2.52. Of course, as seen above in the k-fold specific results, this region has some variability and therefore is more difficult to interpret. It is interpreted as, telomere length reduces by 2.52 if all individuals were exposed to the estimated most severe combination of metals. 

In general, in this NHANES example, we show that in real-world data, \texttt{CVtreeMLE} can answer questions about 1. In a high-dimensional mixed metal exposure space, what is the region that minimizes telomere length? and 2. If we were to force the population to have exposures at those levels, what is the expected outcome compared to our observed population outcomes? In this case, we are not answering what the safe levels are, but what the worse levels are, since a shorter telomere length is indicative of faster biological aging. 

Individual studies have demonstrated that prenatal exposure to specific metals such as barium and molybdenum may impact telomere length in newborns. \cite{Cowell2020} found a significant association where barium contributed to the strongest inverse relationship with newborn telomere length, suggesting that exposure to this metal may accelerate cellular aging processes as indicated by shortened telomeres. Complementarily, \cite{Xia2022UrinaryMetals} reported a positive association between urinary molybdenum and telomere length, particularly in the elderly, which might suggest a protective or mitigating effect of this metal on telomere attrition. These findings are consistent with our results using the CVtreeMLE approach, which not only reaffirms the roles of these metals when analyzed individually but also enhances our understanding of their combined effects on telomere biology. This multi-metal interaction analysis provides a more nuanced view of how exposure to a mixture of metals like barium and molybdenum may influence telomere length and, by extension, long-term health outcomes.

\section{Software}
The CVtreeMLE R package, implements our proposed methodology. It offers comprehensive documentation, including a vignette on semiparametric theory, real-life examples, and comparative analyses. Emphasizing reproducibility, the package includes applications with NIEHS synthetic data and NHANES mixed metal exposure data. Notably, it supports efficient computational performance, suitable for standard personal computers and scalable to high-performance computing for simulations. The package, available on GitHub, exemplifies the push towards accessible, well-documented, and robust statistical software in scientific research.

\section{Discussion}

In this study, we present a novel approach for analyzing mixed exposures, focusing on the data-adaptive identification of critical thresholds in multivariate exposure spaces. Our method introduces a decision tree algorithm integrated with targeted learning to estimate intervention effects within identified exposure regions. This approach leverages cross-validated targeted maximum likelihood estimation to ensure robust and efficient estimation of the attributable regional effect for these regions, making it particularly suitable for complex environmental health data.

Our method is the first to data-adaptively identify safe thresholds in a multivariate exposure context. Traditional approaches often rely on predefined exposure levels or simplistic models that fail to capture the intricate interactions between multiple concurrent exposures in an interpretable way. In contrast, our approach uses a recursive partitioning algorithm to dynamically determine the exposure regions that maximize or minimize health outcomes, addressing the inherent complexities and interactions present in real-world exposure scenarios. This provides policy relevant and interpretable regions in the exposure space that are deemed safe based on our predetermined definition of an oracle \say{safe} region.

Through extensive simulations and application to real-world data, including NIEHS synthetic mixtures and NHANES metal exposure data, we demonstrate the robustness and practical applicability of our method. The simulation studies confirm the asymptotic convergence of our estimator to the true oracle region and its effectiveness in identifying the correct thresholds. The real-world applications further illustrate how our method can identify harmful exposure levels and provide actionable insights for public health interventions.

However, our approach has limitations. When the identified regions differ significantly across folds, it indicates weak signal strength, making the interpretation of the pooled parameter challenging. This variability suggests that the data do not strongly support a single optimal exposure region, complicating regulatory decisions. In such cases, environmental regulators must carefully assess the fold-specific results to understand the potential exposure thresholds better. This limitation underscores the importance of considering the consistency and stability of the identified regions across different data subsets. 

One key aspect of our methodology is the preservation of relative self-selection preferences within the identified regions. This choice aligns with existing literature that often binarizes continuous exposures to facilitate causal inference. By maintaining the natural selection preferences within exposure regions, our approach aims to provide familiar and interpretable estimates of the impact of policy interventions. However, the self-selection strategy may not always represent the most plausible policy intervention. For instance, a policy that moves individuals just below a threshold could result in higher densities near the boundary and lower densities farther from the threshold, reflecting more realistic enforcement scenarios. Expanding our approach to estimate different policies based on domain knowledge and specific regulatory contexts will enhance its applicability and relevance. Future research should explore methods to incorporate such domain-specific policies, providing a more comprehensive toolkit for environmental regulators and policymakers.

Despite these limitations, our method represents a significant advancement in the analysis of mixed exposures. It offers a flexible and powerful tool for uncovering critical exposure thresholds and assessing the combined effects of multiple environmental exposures. The development of the CVtreeMLE R package facilitates the broader application of our methodology, providing researchers and policymakers with a robust, user-friendly tool for evaluating complex exposure data.

In conclusion, our study provides a comprehensive framework for the data-adaptive identification and estimation of safe exposure thresholds in multivariate settings. This approach has substantial implications for public policy, particularly in setting regulatory limits and developing interventions to mitigate the adverse health effects of environmental exposures. Future research should focus on further refining this method, exploring its application to other types of exposure data, and addressing the challenges related to signal variability across folds.

\bibliography{bibliography}

\end{document}